\documentclass[aps,prl,amsmath,amssymb,floatfix,longbibliography,superscriptaddress,twocolumn,PRL]{revtex4-2}

\usepackage{graphicx}
\usepackage{comment}

\begin{document}

\title{Near-Unity-Efficiency Gas Gratings for Ultraviolet, Visible, and Infrared\\ High-Power Lasers}%

\author{K. Ou}
\email[]{ouke025@stanford.edu}
\affiliation{Department of Mechanical Engineering, Stanford University, Stanford, California 94305, USA}

\author{H. Rajesh}
\author{S. Cao}
\author{D. Chakraborty}
\author{V. M. Perez-Ramirez}
\author{D. Singh}
\author{C. Redshaw}
\author{P. Dedeler}
\affiliation{Department of Mechanical Engineering, Stanford University, Stanford, California 94305, USA}

\author{A. Oudin}
\author{E. Kur}
\affiliation{Lawrence Livermore National Laboratory, Livermore, California 94551, USA}

\author{M. M. Wang}
\affiliation{Department of Electrical and Computer Engineering, Princeton University, Princeton, New Jersey 08540, USA}

\author{J. M. Mikhailova}
\affiliation{Department of Mechanical and Aerospace Engineering, Princeton University, Princeton, New Jersey 08540, USA}

\author{L. Lancia}
\affiliation{LULI, CNRS, CEA, Sorbonne Université, École Polytechnique, Institut Polytechnique de Paris, Palaiseau, F-91128, France}

\author{C. Riconda}
\affiliation{LULI, Sorbonne Université, CNRS, École Polytechnique, CEA, Paris, F-75252, France}

\author{P. Michel}
\affiliation{Lawrence Livermore National Laboratory, Livermore, California 94551, USA}

\author{M. R. Edwards}
\email[]{mredwards@stanford.edu}
\affiliation{Department of Mechanical Engineering, Stanford University, Stanford, California 94305, USA}

\begin{abstract}
Interfering deep ultraviolet (DUV) lasers can induce substantial density modulations in an ozone-doped gas flow via photochemical reactions, creating volume diffraction gratings. These transient optics are immune to target debris and shrapnel and feature orders-of-magnitude higher damage thresholds than conventional solid optics, providing a promising method for efficiently manipulating high-energy lasers. In this work, we describe gas gratings that can efficiently diffract probe beams across a variety of wavelengths and pulse durations, ranging from deep ultraviolet to near-infrared and from nanosecond to femtosecond, achieving a full beam diffraction efficiency up to 99\% while preserving the focusability and wavefront quality. In addition, we present a comprehensive characterization of the performance of the gas gratings under various experimental conditions, including imprint fluence, gas composition, and grating geometries, showing significant enhancement of this process with the addition of carbon dioxide. We also demonstrate stable performance over hours of operation. Our results validate a previously developed theoretical model and suggest optimal parameters to efficiently scale gas gratings to high-energy applications.
\end{abstract}

\maketitle

\section{Introduction}\label{sec:intro}
Optical damage has become a bottleneck for high-energy and high-power lasers. Modern lasers have reached impressive scales, from multi-kilojoule pulses for high energy density physics \cite{spaeth2016description, maywar2008omega} to petawatt peak powers for ultrafast science \cite{nees2020zeus, radier202210, yoon2021realization}. These powerful systems have led to breakthroughs in inertial confinement fusion \cite{abu2022lawson, abu2024achievement}, compact particle accelerators \cite{tajima1979laser, gonsalves2019petawatt, miao2022multi}, and laboratory astrophysics \cite{fiuza2020electron, chen2023perspectives}. However, the optics needed to handle these beams have not kept pace. Standard solid optics fail around $1-10\,\mathrm{J/cm^2}$ for nanosecond pulses \cite{ristau2009laser} and $0.1-1\,\mathrm{J/cm^2}$ for sub-picosecond pulses \cite{poole2013femtosecond}, forcing the use of mirrors and lenses that are prohibitively large and costly to manufacture. This limitation becomes even more severe in fusion and other target-facing applications, where the final optics must survive not only the intense laser pulse but also bombardment by debris and particles from target interactions. Current large lasers require meter-scale mirrors that are both expensive and vulnerable to damage accumulation over time; higher-power lasers will require new types of optics.

Gas-based optics are a potential alternative to conventional solid-state optics for the highest energy lasers. Nanosecond laser pulses can propagate through gases at fluences exceeding $1\,\mathrm{kJ/cm^2}$, two-to-three orders-of-magnitude higher than the damage threshold of conventional optics, so a gas optic can be substantially smaller than a solid equivalent. Moreover, gas optics can be created in a continuous gas flow and refreshed at high repetition rate, making them intrinsically immune to accumulated long-term damage and debris. 

However, common gases at standard temperature and pressure have refractive indices that differ from unity by only $\sim\!10^{-4}$, so creating gas-phase optics is challenging. Early refractive gas lenses \cite{durfee1993light, katzir2009plasma, gaul2000production} proved insufficiently flexible for most applications. More recent demonstrations of diffractive gas optics, either laser-written \cite{michine2020ultra} or transducer-driven \cite{schrodel2024acousto}, offer capabilities more suitable for applications. The novel laser-driven scheme proposed by Michine and Yoneda \cite{michine2020ultra} is particularly interesting, since it supports the creation of density variations in a gas larger than $10\%$ and allowed portions of a beam to be diffracted with up to $96\%$ efficiency with a 1-cm-thick optic. The process takes advantage of the efficient energy deposition of a pair of deep ultraviolet (DUV) imprint lasers in an ozone-doped gas flow via photodissociation in the ozone Hartley band \cite{schinke2010photodissociation} and the following chemical reactions that thermalize the released energy in translational modes on a time scale of several nanoseconds \cite{michel2024photochemically, michine2024large, oudin2025piafs}, resulting in a substantial temperature rise at the constructive fringes of the imprint interference pattern. The heating can be treated as local and instantaneous because the spatial features in the interference pattern are much longer than the gas mean free path and the imprint pulse duration is much shorter than the relevant hydrodynamic time scales. As a result, a temperature modulation that matches the imprint interference pattern will be created in the gas. This pure temperature modulation corresponds to the superposition of two counterpropagating acoustic waves and a stationary entropy wave. As the acoustic waves propagate, the temperature modulation will evolve into a density modulation, producing an oscillating grating pattern with a density amplitude reaching a significant fraction of the average density and refractive index modulations on the order of $10^{-5}$ to $10^{-4}$, since the gas refractive index is directly density-dependent. This periodic index structure will then function as a volume transmission grating. 

\begin{figure*}[tb]
    \centering
    \includegraphics[width=\linewidth]{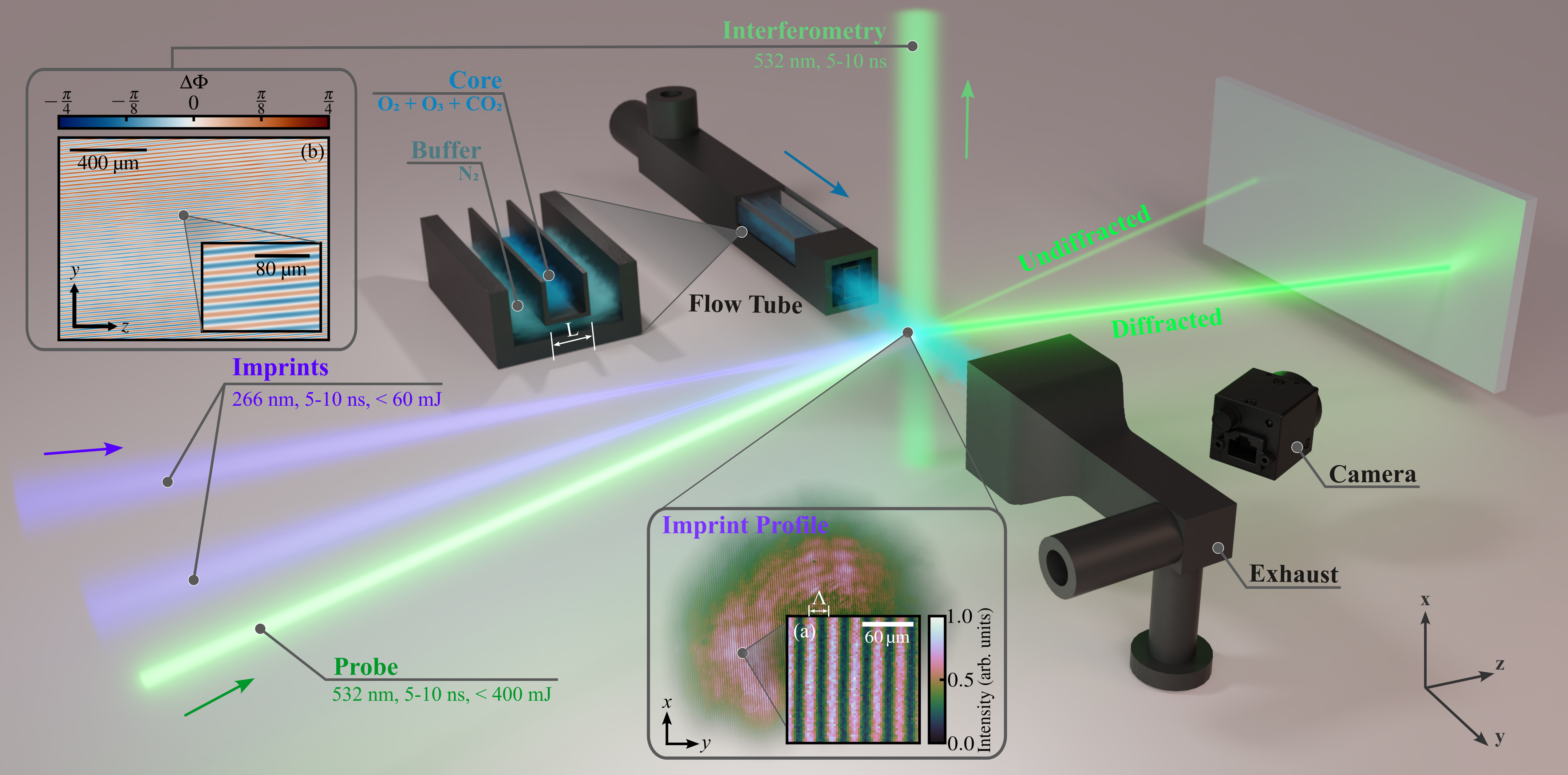}
    \caption{Experimental schematic. Two coherent $266\,\mathrm{nm}$ lasers cross in an ozone-doped flow, producing a large gas density modulation that matches the interference pattern of the imprint lasers. The modulated gas then acts as a volume diffraction grating and deflects a probe beam. The insets show (a) the interference pattern of the imprints and (b) the phase shift due to the gas density modulations from an interferometry measurement.}
    \label{fig:schematic}
\end{figure*}

Although this basic mechanism underlying ozone gas gratings has been demonstrated in proof-of-principle experiments, their suitability as optics for applications has not yet been shown. In this paper, we describe the performance of gas gratings across a wide range of conditions, including demonstrating that they can be used for wavelengths from $266\,\mathrm{nm}$ through $1064\,\mathrm{nm}$ and both nanosecond and femtosecond pulses, with the highest diffraction efficiency above $99\%$. Since our efficiency is measured based on the full beam energy, these results have substantially outperformed previous records \cite{michine2020ultra}. In addition, we present a comprehensive characterization of the grating dynamics and performance as a function of imprint fluence, gas composition, and grating geometries. Our results validate the theoretical model developed by Michel et al. \cite{michel2024photochemically}, showing that the addition of carbon dioxide greatly improves the diffraction efficiency by more than three times. We also measure the grating angular bandwidth at different grating periods and grating thicknesses, demonstrate long-term operation at $10\,\mathrm{Hz}$ over two hours with an average diffraction efficiency above $95\%$, and show that the grating introduces no wavefront distortion in the diffracted beam. These results suggest that gas gratings can now achieve the performance levels required for practical high-energy laser applications. Our systematic characterization provides a roadmap for scaling these transient optics to real-world systems.

\section{Theory}\label{sec:theory}
Here we summarize the key equations characterizing the diffraction efficiency of a volume diffraction grating in the Bragg regime. These relations can be derived from the coupled mode theory \cite{yeh1993introduction} and will aid in interpreting the experimental results in Section \ref{sec:results}. Consider a grating with spatial period $\Lambda$ and thickness $D$. In experiments, $\Lambda$ is the period of the imprint interference pattern and can be controlled by the crossing angle between the two imprint beams. A monochromatic probe beam with wavelength $\lambda$ will have its first-order diffraction efficiency $\eta_1$ maximized if the incident angle is the Bragg angle $\theta_B$, defined as:
\begin{equation}
    \theta_B = \arcsin{\left(\frac{\lambda}{2\Lambda}\right)}
    \label{eq:theta_B}
\end{equation}
In this case, the diffraction efficiency is \cite{yeh1993introduction}:
\begin{equation}
    \eta_1 = \sin^2{(\kappa D)}
    \label{eq:eta}
\end{equation}
where $\kappa = \pi n_1 / (\lambda \cos\theta_B)$ is the coupling coefficient and $n_1$ is the index modulation (i.e. assuming the refractive index is $n(x) = n_0 + n_1\cos{(Kx)}$ where $n_0$ is the background refractive index, $K = 2\pi/\Lambda$ is the grating wavenumber, and $x$ is the spatial coordinate). A diffraction efficiency of $100\%$ can be theoretically achieved if $\kappa D = \pi / 2$, yielding
\begin{equation}
    \frac{2n_1D}{\lambda} \approx 1
    \label{eq:optimal_cond}
\end{equation}
if $\theta_B$ is small. In practice, a diffraction efficiency of unity cannot be achieved because any physical beam has a finite angular and spectral bandwidth, so it is impossible to have the entire beam incident at the Bragg angle. When such a phase mismatch is present, the diffraction efficiency becomes \cite{yeh1993introduction}:
\begin{equation}
    \eta_1 = \frac{\kappa^2}{\kappa^2 + \left(\Delta\alpha/2\right)^2}\sin^2{\left\{\kappa D\left[1 + \left(\frac{\Delta\alpha}{2\kappa}\right)^2\right]^{1/2}\right\}},
    \label{eq:eq_eta}
\end{equation}
where
\begin{equation}
    \kappa^2 = \left(\frac{\pi n_1}{\lambda}\right)^2 \frac{1}{\cos{\theta_1}\,\cos{\theta_2}}
    \label{eq:eq_kappa}.
\end{equation}
Here $\theta_1 = -\theta_B + \Delta\theta$ is the incident angle and $\theta_2 = \theta_B + \Delta\theta$ is the first-order diffracted angle for angular mismatch $\Delta\theta$ and phase mismatch $\Delta\alpha = -K\,\Delta\theta$. 

\section{Methods}\label{sec:methods}
A schematic of the experiments described here is shown in Fig. \ref{fig:schematic}. A $10\,\mathrm{Hz}$ Q-switched Nd:YAG laser with two beamlines (SpectraPhysics PIV-400) produced the imprint beams and probe beam. One of the beamlines was frequency-quadrupled using temperature controlled DKDP and KDP crystals. The generated $266\,\mathrm{nm}$ pulses contained up to $60\,\mathrm{mJ}$ of energy and were split to form the two imprint beams. The imprints crossed in an ozone-doped gas flow with adjustable crossing angle. The path lengths of the two imprint beams were matched to ensure temporal coherence and good contrast in the interference pattern. The other beamline provided the probe and could be frequency-doubled or quadrupled. An electronic delay generator controlled the time delay between the imprint pulses and the pulsed probe. The full width at half maximum (FWHM) pulse duration for all beams was $5-10\,\mathrm{ns}$. We imaged these beams at the gas to estimate the local imprint fluence, as shown in Fig. \ref{fig:schematic}(a). In addition, a $532\,\mathrm{nm}$ pulse was sampled from the probe to measure the refractive index modulation in the gas via Mach-Zehnder interferometry. As shown in Fig. \ref{fig:schematic}(b), a substantial index modulation can be created in the gas, corresponding to a $> 10\%$ density modulation.

A continuous-wave (CW) $266\,\mathrm{nm}$ beam (not shown in Fig. \ref{fig:schematic}) was used to monitor the ozone concentration via its measured absorption and the Beer-Lambert law. A CW $532\,\mathrm{nm}$ beam (not shown in Fig. \ref{fig:schematic}, but following a similar path as the pulsed probe) was diffracted by the gas grating onto a photodetector, providing a single-shot measurement of time-resolved grating dynamics.

Gas gratings with varying thickness were created using 3D-printed flow tubes with varying cross-section dimensions. These flow tubes consisted of two channels: an inner channel for the ozone-doped core flow and an outer channel for a nitrogen buffer flow. By matching the buffer and core flow velocities, we were able to avoid a shear layer and resultant turbulent mixing at the ozone interface, which would cause the ozone to rapidly diffuse. The performance of the optic is not strongly dependent on mixing at the nitrogen-air interface, and this coflow configuration offers superior performance to a single-channel tube. The ozone was created using a corona discharge ozone generator (Absolute Ozone ATLAS 30C), reaching concentrations up to $6\%$. All ozone concentrations were measured using the CW UV probe under the assumed flow conditions of $1\,\mathrm{atm}$, $300\,\mathrm{K}$. 

\begin{figure*}[tb]
    \centering
    \includegraphics[width=\linewidth]{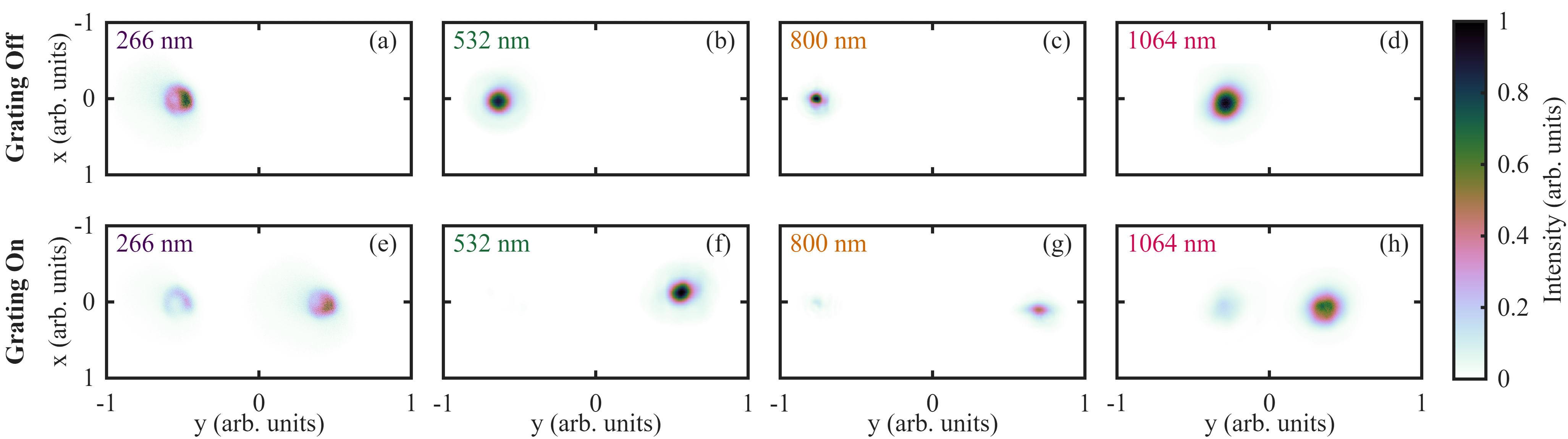}
    \caption{Images of transmitted pulsed probes of various wavelengths with and without the presence of gas gratings. The diffraction efficiency is $57\%$ for the $266\,\mathrm{nm}$ probe ($84\%$ was transmitted and $68\%$ of the transmitted was diffracted to the first order), $99\%$ for the $532\,\mathrm{nm}$ probe, $85\%$ for the $800\,\mathrm{nm}$ probe, and $81\%$ for the $1064\,\mathrm{nm}$ probe. The $800\,\mathrm{nm}$ probe has a pulse duration of $\sim 35\,\mathrm{fs}$ while the rest have pulse durations of $\sim 5-10\,\mathrm{ns}$. The spectrum of the femtosecond probe spans from $795\,\mathrm{nm}$ to $815\,\mathrm{nm}$ (FWHM). The images were averaged of 100 shots.}
    \label{fig:diffraction}
\end{figure*}

The primary metric for the performance of the gas gratings is the diffraction efficiency of the pulsed probe, defined as the fraction of the incident probe energy that is diffracted into the first order. For many measurements presented here, this was approximated by comparing the diffracted and undiffracted transmitted beams imaged on a Teflon scattering screen or directly on a CMOS chip, since energy absorbed, scattered, or diffracted to higher orders was negligibly small ($\ll 1\%$) for all probe wavelengths except $266\,\mathrm{nm}$. The primary parameter scans were performed using a $532\,\mathrm{nm}$ nanosecond probe.

\section{Results}\label{sec:results}
We tested gas gratings across a range of probe wavelengths and durations; Figure \ref{fig:diffraction} shows efficient diffraction of $266\,\mathrm{nm}$, $532\,\mathrm{nm}$, and $1064\,\mathrm{nm}$ nanosecond pulses and $800\,\mathrm{nm}$ 35-femtosecond pulses. The highest diffraction efficiency that can be achieved depends on both the angular bandwidth of the gratings and the spatial quality of the probe, since only the angular components that fall within the grating angular bandwidth will be diffracted. The $532\,\mathrm{nm}$ probe was cleaned using a spatial filter, resulting in a diffraction efficiency of $99\%$. This suggests that lower peak efficiencies observed under other conditions can largely be attributed to poor probe beam quality; there is no reason to expect that similarly high efficiencies cannot be achieved for all wavelengths. Usefully, the reduction in efficiency due to poor incident beam quality corresponds to an increase in the diffracted beam quality.

Perhaps surprisingly, it is also possible to efficiently diffract a beam at $266\,\mathrm{nm}$, as shown in Fig. \ref{fig:diffraction}(a), despite the fact that the concentration of ozone in the grating is sufficient to almost entirely absorb ultraviolet light over a distance of the grating thickness. Here, we are operating the probe well above the saturation fluence of ozone ($\sim\!77\,\mathrm{mJ/cm^2}$ at $266\,\mathrm{nm}$ \cite{daumont1992ozone}), a condition readily achieved for the high-energy beams of interest for these optics, so that only a small fraction of the probe is absorbed. Any hydrodynamic response to the energy deposited by the probe does not appear until the beam has already left the grating. 

The example images shown in Fig. \ref{fig:diffraction} suggest that the diffraction efficiencies of gas gratings are sufficient to be useful for applications for a wide range of laser types. In the following sections, we present a more detailed characterization of these gas gratings under various experimental conditions. Specifically, in Section \ref{sec:dynamics}, we examine the temporal dynamics of the gas gratings; in Section \ref{sec:energy}, we characterize the grating performance at various imprint fluence and gas composition; in Section \ref{sec:bandwidth}, we measure the angular bandwidth of the grating as a function of the grating length and grating period; and finally, in Section \ref{sec:application}, we study features of the gas gratings relevant for applications, including the spatial quality of the diffracted probe and the long-term stability.

\subsection{Grating Dynamics}\label{sec:dynamics}
Since the imprint pulses are much shorter than the hydrodynamic time scale of the gas, they instantaneously imprint a temperature modulation in the gas. Consequently, two counterpropagating acoustic waves and a stationary entropy wave are launched. The resultant density—and therefore refractive index modulation—are created by the superposition of these three modes. We can clearly resolve the temporal dynamics using both traces of the CW probe and delay scans of the nanosecond probe.

\begin{figure}[tb]
    \centering
    \includegraphics[width=\linewidth]{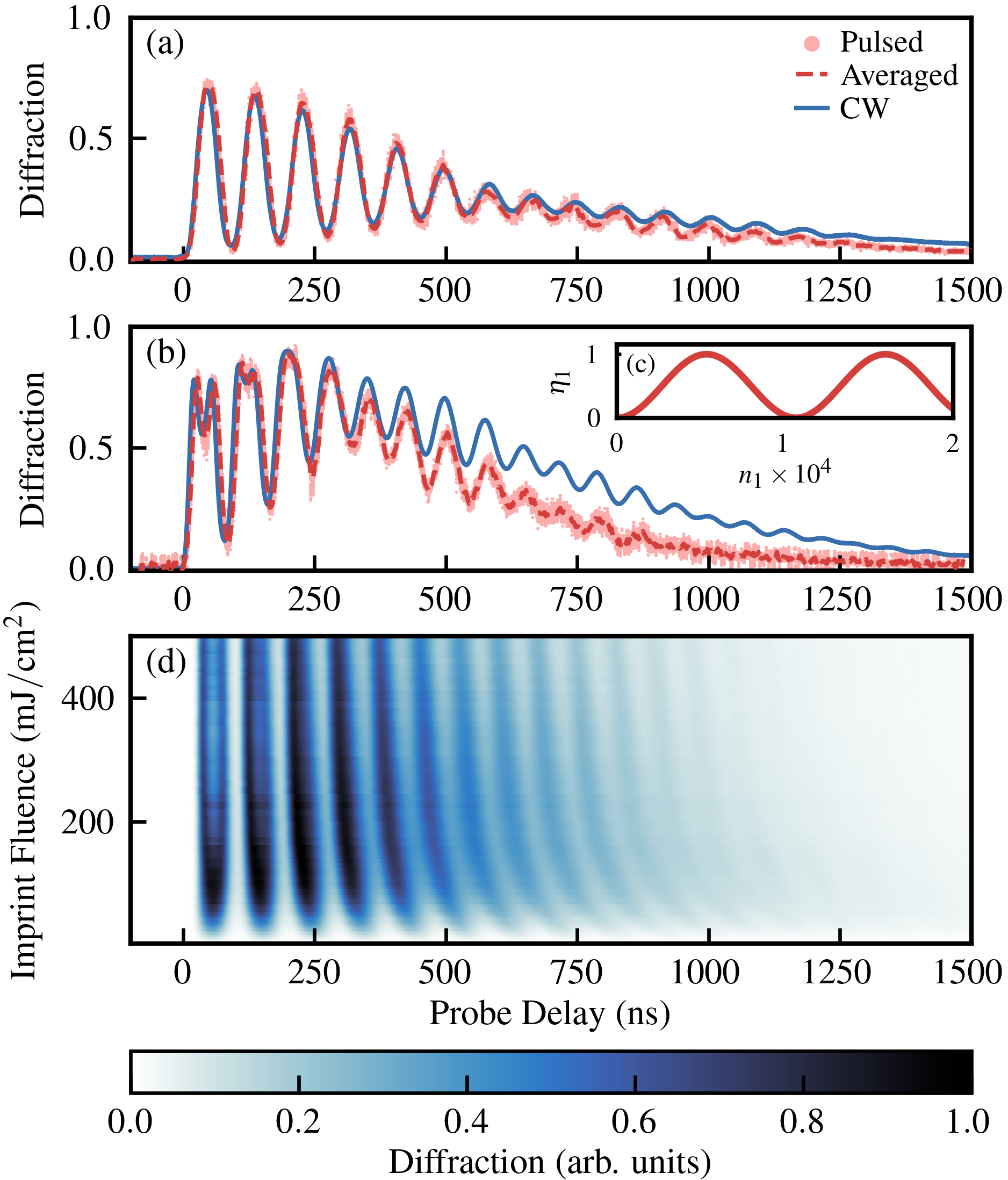}
\caption{(a,b) Comparison of the $532\,\mathrm{nm}$ nanosecond probe delay scans (red dots/curves: single-shot/averaged measurements) with normalized CW probe diffraction traces (blue curves) under two different conditions: the imprint fluence in (b) was roughly double that of (a) while the ozone concentration was around $2.3\%$ in both measurements. (c) First-order diffraction efficiency $\eta_1$ vs. index modulation $n_1$ according to \eqref{eq:eta}, with $D = 5\,\mathrm{mm}$, $\theta_B = 0.5^\circ$, and $\lambda = 532\,\mathrm{nm}$. (d) CW traces at varying imprint fluences with $3.9\%\pm0.1\%$ $\mathrm{O_3}$ (assuming $1\,\mathrm{atm},\,300\,\mathrm{K}$), $5\,\mathrm{mm}$ wide flow tube, and $\sim\!32\,\text{\textmu}\mathrm{m}$ grating period.}
    \label{fig:delay_scan}
\end{figure}

Figure \ref{fig:delay_scan}(a) and (b) show the temporal dynamics of the gas gratings at two distinct conditions: the imprint fluence in (b) is approximately twice that used in (a), while the ozone concentration was similar. The fast oscillations originate from the standing wave formed by the two counterpropagating acoustic modes. The oscillation period $\tau$ can be estimated as $\tau = \Lambda / c_s$ where $\Lambda$ is the grating period and $c_s$ is the average speed of sound in the gas. Since $c_s$ is a function of gas temperature and the density modulation is directly driven by the temperature modulation, we can estimate the average gas temperature using these CW traces, as detailed in Section \ref{sec:energy}. (This is the same technique used to measure temperature and flow velocity in flames and gases; see Ref. \cite{eichler2013laser} and references therein. However, here, unlike in previous diagnostic development, the energy deposited by the imprint beams substantially raises the average temperature.)

The traces also reveal an underlying envelope in which the diffraction efficiency remains above zero even at the oscillation troughs, indicating a persistent gas density modulation as the acoustic modes propagate. This phenomenon is also observed in numerical simulations \cite{oudin2025piafs}, where theory suggests that the background density modulation arises from the different damping rates of the acoustic modes and the entropy mode. When the density modulation is large, profile steepening occurs because of the varying group velocities in high-density and low-density regions, resulting in energy transfer to higher-order harmonics. Since the damping rate of an acoustic wave increases proportional to the square of the wavenumber $K$ \cite{pierce2019acoustics}, energy rapidly dissipates through the higher-order harmonics. The entropy mode, however, is non-propagating, so its damping is dominated by thermal diffusion, which is much slower. The persistent entropy mode produces the non-zero diffraction efficiencies at the oscillation troughs. 

Finally, as shown in the first two periods in Fig. \ref{fig:delay_scan}(b), double-peak structures arise when the index modulation is too high. This can be explained from \eqref{eq:eta}, noting that the diffraction efficiency for a volume transmission grating starts to drop if $\kappa D$ rises above $\pi / 2$, where $\kappa\propto n_1$. 

In both cases, the CW traces and pulsed probe delay scans show similar temporal features, where small discrepancies likely arise from differences in the spatial and temporal profiles of the two probes and the spatial non-uniformity of the gratings. Figure \ref{fig:delay_scan}(d) illustrates the parameter space probed by the CW beam at a particular gas mixture and grating geometry. Changing the latter two parameters should only rescale the parameter space without fundamentally distorting its overall shape. This provides valuable guidance for selecting optimal parameters that produce maximum diffraction efficiency with minimal energy in the imprint beams, which will be detailed in the next section.

\subsection{Energy Deposition}\label{sec:energy}
\begin{figure}[tb]
    \centering
    \includegraphics[width=\linewidth]{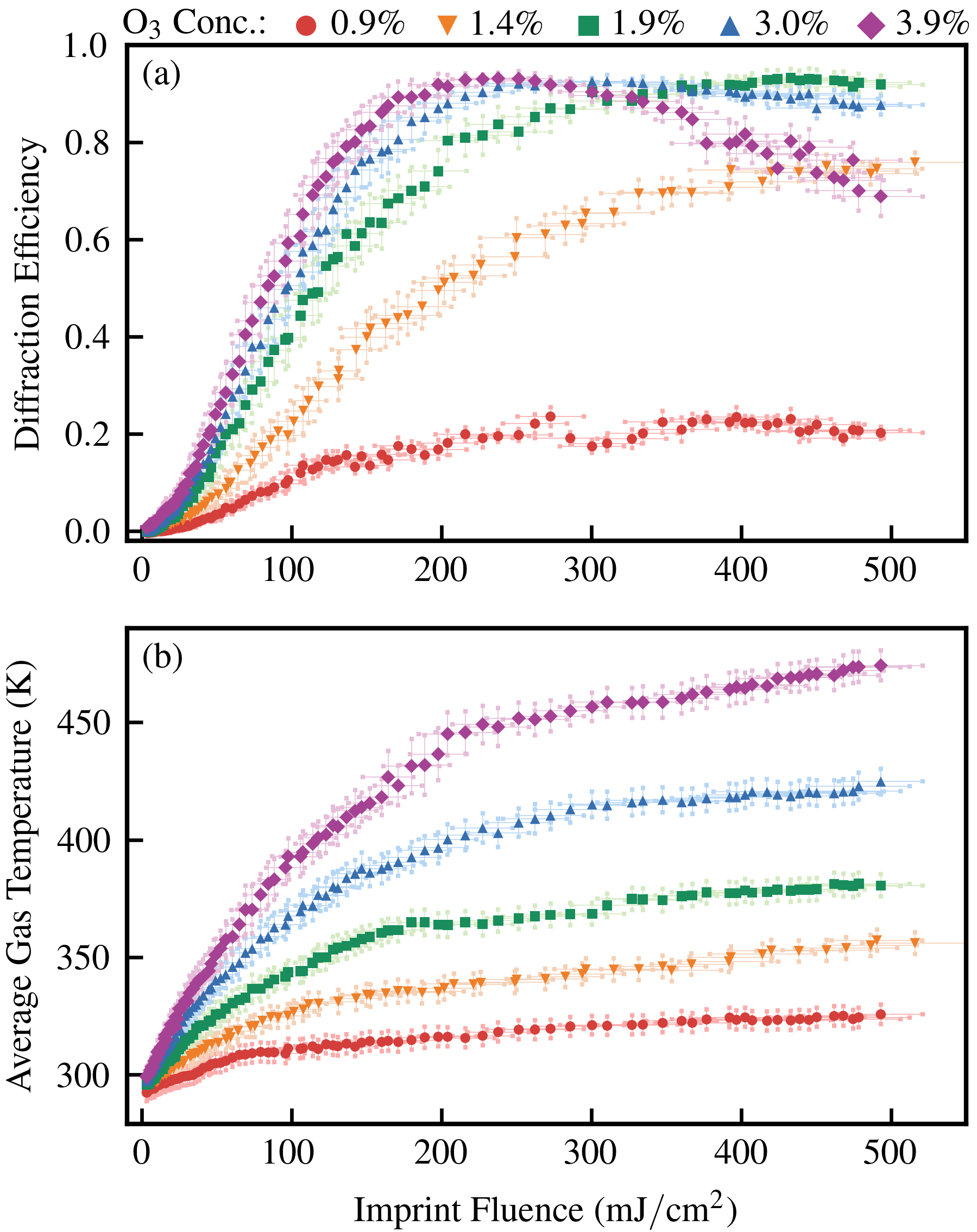}
    \caption{(a) Diffraction efficiency of the $532\,\mathrm{nm}$ nanosecond probe ($\approx\!5\,\mathrm{ns}$ FWHM pulse duration, $\lessapprox\!200\,\text{\textmu}\mathrm{J}$ average pulse energy, cleaned with a spatial filter) and (b) average gas temperature estimated from CW traces versus imprint fluences at different ozone concentrations ($0.9\%\pm0.1\%,\,1.4\%\pm0.1\%,\,1.9\%\pm0.2\%,\,3.0\%\pm0.1\%,\,3.9\%\pm0.1\%$, indicated by colors). The ozone concentration is calculated assuming $1\,\mathrm{atm},\,300\,\mathrm{K}$. Error bars indicate uncertainty about the local imprint fluence and standard deviations of the measured efficiency. The flow tube was $5\,\mathrm{mm}$ wide. Grating period was around $32\,\text{\textmu}\mathrm{m}$.}
    \label{fig:energy_scan}
    \vspace{-10pt}
\end{figure}

The peak amplitude of the refractive index modulation $n_1$ depends on the imprint fluence, absorber (ozone) density, and gas composition, as these parameters determine the amount of energy deposited into the gas. Figure \ref{fig:energy_scan} shows the diffraction efficiency of a $532\,\mathrm{nm}$ nanosecond probe at various imprint fluences and ozone concentrations. The efficiency was measured when the index modulations peaked in time based on the CW traces (i.e. the first oscillation peak in Fig. \ref{fig:delay_scan}(a) or the first dip in Fig. \ref{fig:delay_scan}(b) when double-peak structures were present). We varied the incident imprint energy to adjust the fluence. The ozone concentration was controlled by varying the power supplied to the ozone generator. Although the pulse duration of the imprint beams was slightly longer at low energy, it remained much shorter than the acoustic periods of the response, so the energy deposition can still be approximated as instantaneous.

As shown in Fig. \ref{fig:energy_scan}(a), the diffraction efficiency in general increases as the gas is pumped harder, until the index modulation exceeds the value for optimal diffraction (i.e. $\kappa D > \pi / 2$; cf. Fig. \ref{fig:delay_scan}(c)). At low imprint fluence, the efficiency is similar across different ozone concentrations as the imprint beams are depleted. Conversely, at high imprint fluence and low ozone concentration, the efficiency saturates because of ozone depletion, preventing additional energy deposition into the gas. When both the imprint fluence and the ozone concentration are high, the efficiency may start to drop as the index modulation exceeds the optimum. In this case, we can achieve a higher diffraction efficiency by adjusting the probe delay so that $n_1$ is lower. However, as will be discussed in Section \ref{sec:application}, operating near the efficiency peak provides better stability. Alternatively, if the path lengths of the two imprint beams are not well aligned so that the contrast of the imprint interference pattern is not ideal, we will observe a similar decrease in diffraction efficiency at high imprint fluences, because the destructive interference fringes will be heated while the absorption at the constructive interference fringes is saturated, so the overall temperature modulation becomes weaker. Figure \ref{fig:energy_scan}(b) shows the average gas temperature as a function of the imprint fluences and ozone concentration. The temperature was estimated from the relationship between the gas temperature and speed of sound ($T = M c_s^2 / \gamma R$), where $R$ is the universal gas constant, $\gamma$ and $M$ are the average heat capacity ratio and molar mass of the gas mixture, respectively, and $c_s = \Lambda / \tau$ is the average speed of sound, which can be estimated from the CW traces as in Fig. \ref{fig:delay_scan}. The increase in average gas temperature at higher imprint fluence or ozone concentration is a signature of more effective energy deposition.

The gas mixture used for the experimental results shown in Fig. \ref{fig:delay_scan} and Fig. \ref{fig:energy_scan} contained only $\mathrm{O}_2$ and $\mathrm{O}_3$. Adding $\mathrm{CO_2}$ has been predicted to dramatically enhanced the index modulation \cite{michel2024photochemically} through two complementary mechanisms. First, carbon dioxide has a higher refractive index than oxygen ($1 + 4.5\times10^{-4}$ for $\mathrm{CO_2}$ vs. $1 + 2.7\times 10^{-4}$ for $\mathrm{O_2}$ at standard temperature and pressure), so a $\mathrm{CO}_2$-enriched mixture will have a higher index modulation than a $\mathrm{CO}_2$-free mixture with the same density modulation amplitude. Second, the addition of $\mathrm{CO}_2$ allows for faster quenching because of the reaction \cite{michel2024photochemically}
\begin{equation}
    \mathrm{O}(^1D) + \mathrm{CO_2} \xrightarrow{} \mathrm{O} + \mathrm{CO_2},
    \label{eq:CO2}
\end{equation}
so that more heat can be released within the relevant time scale. 

\begin{figure}[tb]
    \centering
    \includegraphics[width=\linewidth]{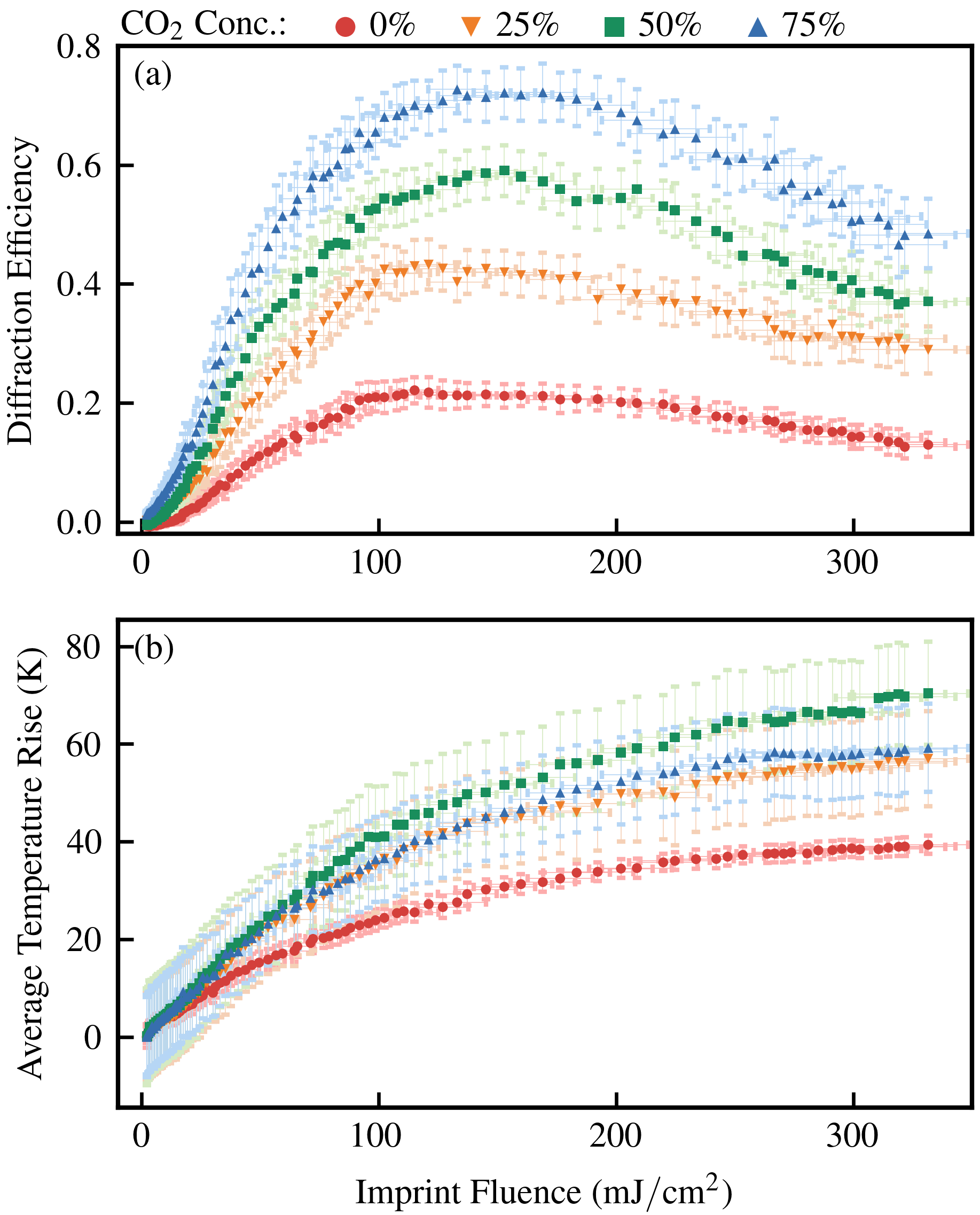}
    \caption{(a) Diffraction efficiency of the $532\,\mathrm{nm}$ nanosecond probe ($\approx\!5\,\mathrm{ns}$ FWHM pulse duration, $\lessapprox\!200\,\text{\textmu}\mathrm{J}$ average pulse energy, cleaned with a spatial filter) and (b) average gas temperature rise estimated from CW traces versus imprint fluence at different $\mathrm{CO_2}$ concentrations ($0\%$, $25\%\pm5\%$, $50\%\pm5\%$, and $75\%\pm5\%$, indicated by colors). The $\mathrm{CO_2}$ concentration are estimated as the ratio of $\mathrm{CO_2}$ flow rate to the total $\mathrm{CO_2}$-$\mathrm{O_2}$-$\mathrm{O_3}$ mixture flow rate, with $\mathrm{CO_2}$ and $\mathrm{O_2}$ initially at similar pressure and temperature. Ozone concentration remains constant at $1.0\%\pm0.1\%$ across all measurements (calculated assuming $1\,\mathrm{atm}$, $300\,\mathrm{K}$). Error bars indicate uncertainty about the local imprint fluence and standard deviations of the measured efficiency. The flow tube was $5\,\mathrm{mm}$ wide. Grating period was around $30\,\text{\textmu}\mathrm{m}$.}
    \label{fig:CO2_scan}
\end{figure}
As shown in Fig. \ref{fig:CO2_scan}, we examined the effects of mixing $\mathrm{CO_2}$ into the gas. The ozone concentration was fixed at $1\%\pm0.1\%$. The $\mathrm{CO_2}$ concentration was estimated as the ratio of the $\mathrm{CO_2}$ flow rate to the total mixture flow rate, with all gases maintained at approximately equal temperature and pressure. The probe delay was again set to the index modulation peaks according to the CW traces. 

The addition of up to $\sim\!75\%$ $\mathrm{CO_2}$ to the mixture resulted in a more than three-fold enhancement in the index modulation and the corresponding diffraction efficiency. This $\mathrm{CO_2}$-enriched mixture offers significant advantages in experimental environments where achieving high $\mathrm{O_3}$ concentrations is difficult, particularly in vacuum systems where increased $\mathrm{O_3}$ dissociation during pressurization is unavoidable. As shown in Fig. \ref{fig:CO2_scan}(b), the average gas temperature increases at higher imprint fluences due to enhanced energy deposition. 

\subsection{Grating Bandwidth}\label{sec:bandwidth}
\begin{figure}[tb]
    \centering
    \includegraphics[width=\linewidth]{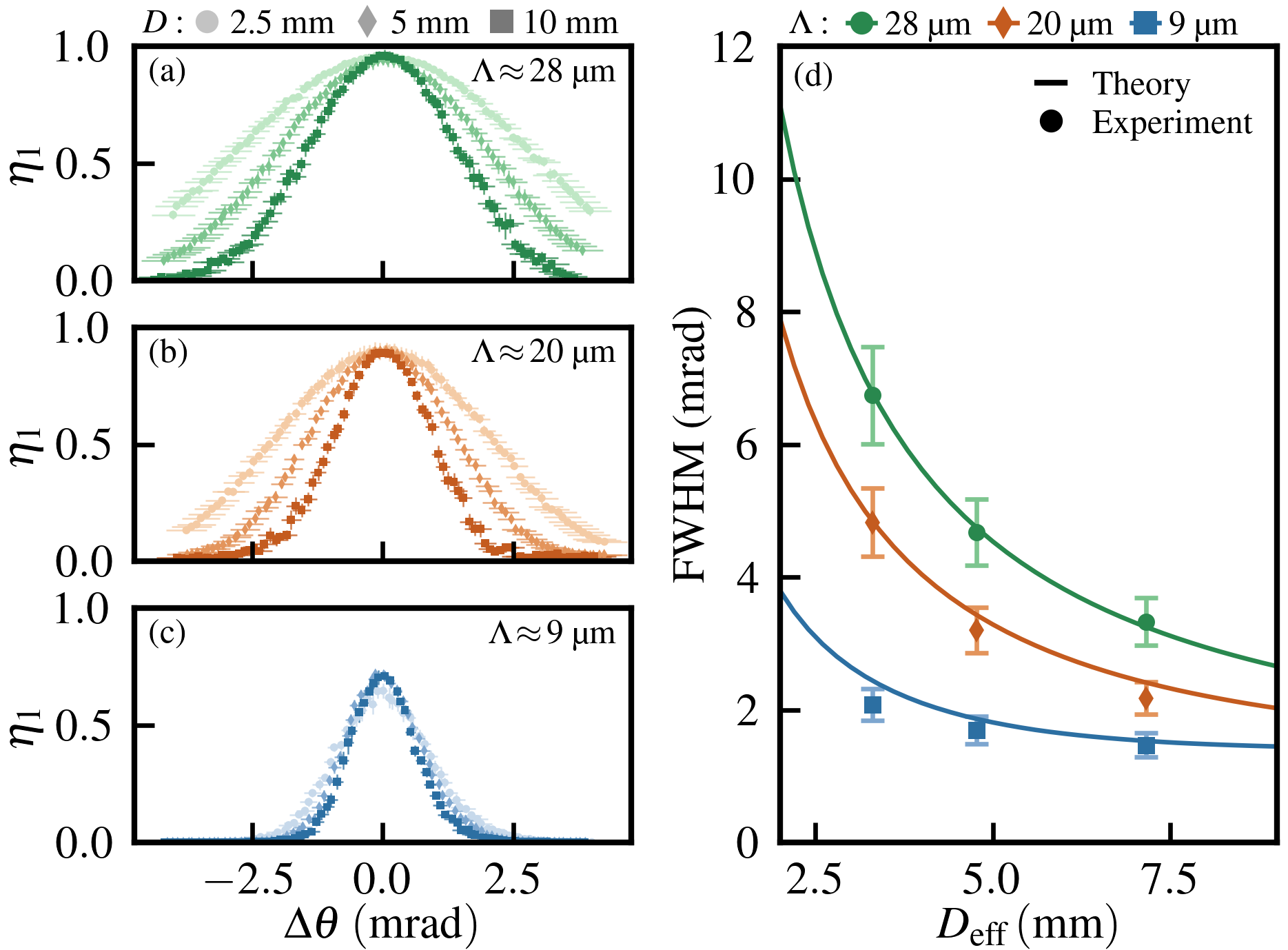}
    \caption{(a-c) Diffraction efficiency of the $532\,\mathrm{nm}$ nanosecond probe beam versus angular deviation from the Bragg angle $\Delta\theta$ for $2.5\,\mathrm{mm}$-, $5\,\mathrm{mm}$-, and $10\,\mathrm{mm}$-thick flow tubes (indicated by different markers/shades) and three grating periods: (a) $28\,\text{\textmu}\mathrm{m}$, (b) $20\,\text{\textmu}\mathrm{m}$, and (c) $9\,\text{\textmu}\mathrm{m}$ (indicated by different colors). The imprint energy, probe delay, and gas composition were optimized for maximum diffraction efficiency in each configuration. (d) FWHM of the envelopes in (a-c) vs. effective grating thickness at various grating periods (markers and error bars) and comparison to theory according to \eqref{eq:eq_eta_total} (solid lines). The effective grating thickness was fit to minimize differences in the envelopes between experimental measurements in (a-c) and theoretical predictions according to \eqref{eq:eq_eta_total}. The fitted thickness were $3.30\,\mathrm{mm}$, $4.77\,\mathrm{mm}$, $7.16\,\mathrm{mm}$ for $2.5\,\mathrm{mm}$-, $5\,\mathrm{mm}$-, and $10\,\mathrm{mm}$-thick flow tubes, respectively. The divergence angle was $\theta_d \approx 1\,\mathrm{mrad}$. For theory, the index modulations were chosen as $n_1 = \pi\lambda / (2D)$ to maximize the diffraction efficiency.}
    \label{fig:bandwidth}
\end{figure}

The angular bandwidth of the gas gratings depends on the grating period $\Lambda$ and grating thickness $D$. The half bandwidth is defined as the angular deviation from the Bragg angle $\theta_B$ that drops the efficiency to $50\%$. Figure \ref{fig:bandwidth}(a-c) illustrate the diffraction efficiency of the $532\,\mathrm{nm}$ probe versus the incident angle with various flow tube dimensions and grating periods. Since the probe beam has a non-negligible divergence, the diffraction efficiency is the convolution of the grating and probe angular spectra, as shown in Fig. \ref{fig:bandwidth}(a-c). In experiments, the grating period was adjusted by changing the crossing angle of the imprint beams, and the grating thickness was varied by using flow tubes with different cross-section dimensions. The effective grating thickness, however, may differ from the flow tube widths. For thicker gratings, depletion of the imprint lasers will reduce the effective thickness. For thinner gratings, diffusion of ozone at the grating surfaces can noticeably increase the effective thickness.

To compare our measurements to theory, recall the diffraction efficiency $\eta_1$ of a plane wave through a uniform volume transmission grating as given in \eqref{eq:eq_eta}. A monochromatic laser beam with a Gaussian angular spectrum has a normalized intensity distribution of:
\begin{equation}
    I = \sqrt{\frac{2}{\pi k_d^2}} \exp{\left\{-\frac{2(k-k_i)^2}{k_d^2}\right\}},
    \label{eq:eq_gauss}
\end{equation}
where $k_i = k_0\sin{\theta_i}$ and $k_d = k_0\sin{\theta_d}$ for central incident angle $\theta_i$ and divergence half angle $\theta_d$, $k$ is the wavenumber, and $k_0 = 2\pi/\lambda$ is the laser wavenumber. The observed diffraction efficiency of such a laser pulse through the grating given by \eqref{eq:eq_eta} is:
\begin{equation}
    \eta_{1,\,\mathrm{total}} = \int_{-k_0}^{k_0} I(k)\cdot\eta_1(k)\,dk
    \label{eq:eq_eta_total}
\end{equation}
where $\Delta\theta = \arcsin{(k/k_0)} - \theta_B$ for $\eta_1(k)$ from \eqref{eq:eq_eta}.

The solid lines in Fig. \ref{fig:bandwidth}d show the FWHM of the efficiency vs. angular mismatch envelopes as a function of grating thickness at different grating periods according to \eqref{eq:eq_eta_total}. The markers indicate FWHM of the measured envelopes shown in Fig. \ref{fig:bandwidth}(a-c). To find the relevant parameters, we fit $n_1$, $D$, and $\theta_d$ so that envelopes from the experimental measurements in Fig. \ref{fig:bandwidth}(a-c) can be reproduced in theory, subject to the constraint that the beam divergence remains constant at each grating period. Since the maximum diffraction efficiency we obtained experimentally was close to unity, we chose $n_1$ that satisfies \eqref{eq:optimal_cond} when evaluating \eqref{eq:eq_eta_total}. This gives an effective grating thickness $D_{\mathrm{eff}}$ that differs from the actual width of the flow tubes, which can be explained by gas expansion in air ($D_{\mathrm{eff}} \approx 3.3\,\mathrm{mm}$ for the $2.5\,\mathrm{mm}$-thick flow tube) and imprint depletion($D_{\mathrm{eff}} \approx 7.2\,\mathrm{mm}$ for the $10\,\mathrm{mm}$-thick flow tube).  

In our configurations with close to $100\%$ diffraction efficiency, the grating period is around $30\,\text{\textmu}\mathrm{m}$, corresponding to a $0.25$ degree imprint crossing half angle and a $0.5$ degree Bragg angle for a $532\,\mathrm{nm}$ probe beam, yielding a $1$ degree separation between the diffracted and undiffracted probes. Larger angular separations require shorter grating periods, but this reduces angular bandwidth as shown in Fig. \ref{fig:bandwidth}. To maintain a high diffraction efficiency in this case, the probe divergence must remain much smaller than the grating angular bandwidth. Alternatively, gases with higher refractive indices (e.g. $\mathrm{SF}_6$) or at higher pressure can be used to achieve a higher index modulation, since the bandwidth depends on $n_1$. Meanwhile, although the diffraction efficiency becomes lower with a larger diffraction angle, the diffracted beam will have a better spatial quality as the undesired spatial components will be blocked out of the grating angular bandwidth.

\subsection{Stability and Spatial Quality}\label{sec:application}
\begin{figure}[tb]
    \centering
    \includegraphics[width=\linewidth]{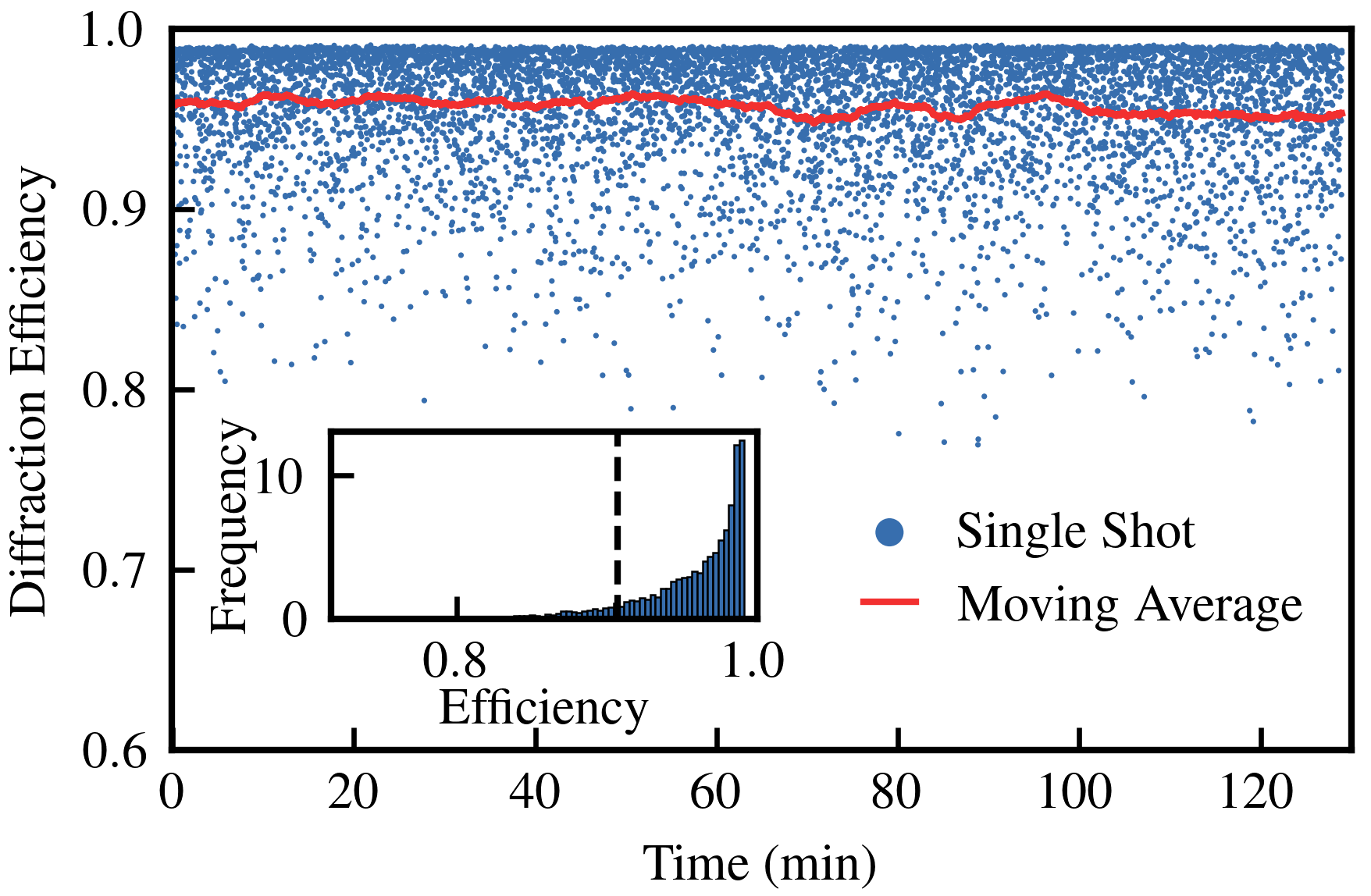}
    \caption{Diffraction efficiency of the $532\,\mathrm{nm}$ nanosecond probe beam measured over two hours. The diffraction efficiency was $95.7\% \pm 3.6\%$ (mean $\pm$ std). Individual blue markers represent single-shot measurements while the red trace shows the moving average calculated over a 5-minute window. The inset histogram shows the distribution of single-shot efficiencies. The vertical dashed line marks the diffraction efficiency threshold exceeded by 90\% of the shots. The grating period was around $30\,\mathrm{\text{\textmu} m}$.}
    \label{fig:stability}
\end{figure}
Results in Fig. \ref{fig:stability} demonstrate the continuous operation of gas gratings at $10\,\mathrm{Hz}$ for more than two hours, achieving an average efficiency above $95\%$ over 7737 shots with no significant long-term degradation. The inset histogram shows the percentage of shots at different efficiency, with $90\%$ of the shots at an efficiency above $90\%$. Operating near the local efficiency maxima as in Fig. \ref{fig:delay_scan}(c) can further stabilize the performance. Note that $\kappa D \approx \pi / 2$ at the first efficiency peak where $\kappa \propto n_1$. Hence, for any small disturbance $\delta\phi$,
\begin{equation}
    \eta_1 = \sin^2{\left(\frac{\pi}{2} + \delta\phi\right)} = \cos^2{(\delta\phi)} \approx 1 - (\delta\phi)^2,
    \label{eq:eq_stability}
\end{equation}
so the diffraction efficiency near the peak is less sensitive to fluctuations in the imprint energy, gas composition, and time jitter.

\begin{figure}[tb]
    \centering
    \includegraphics[width=\linewidth]{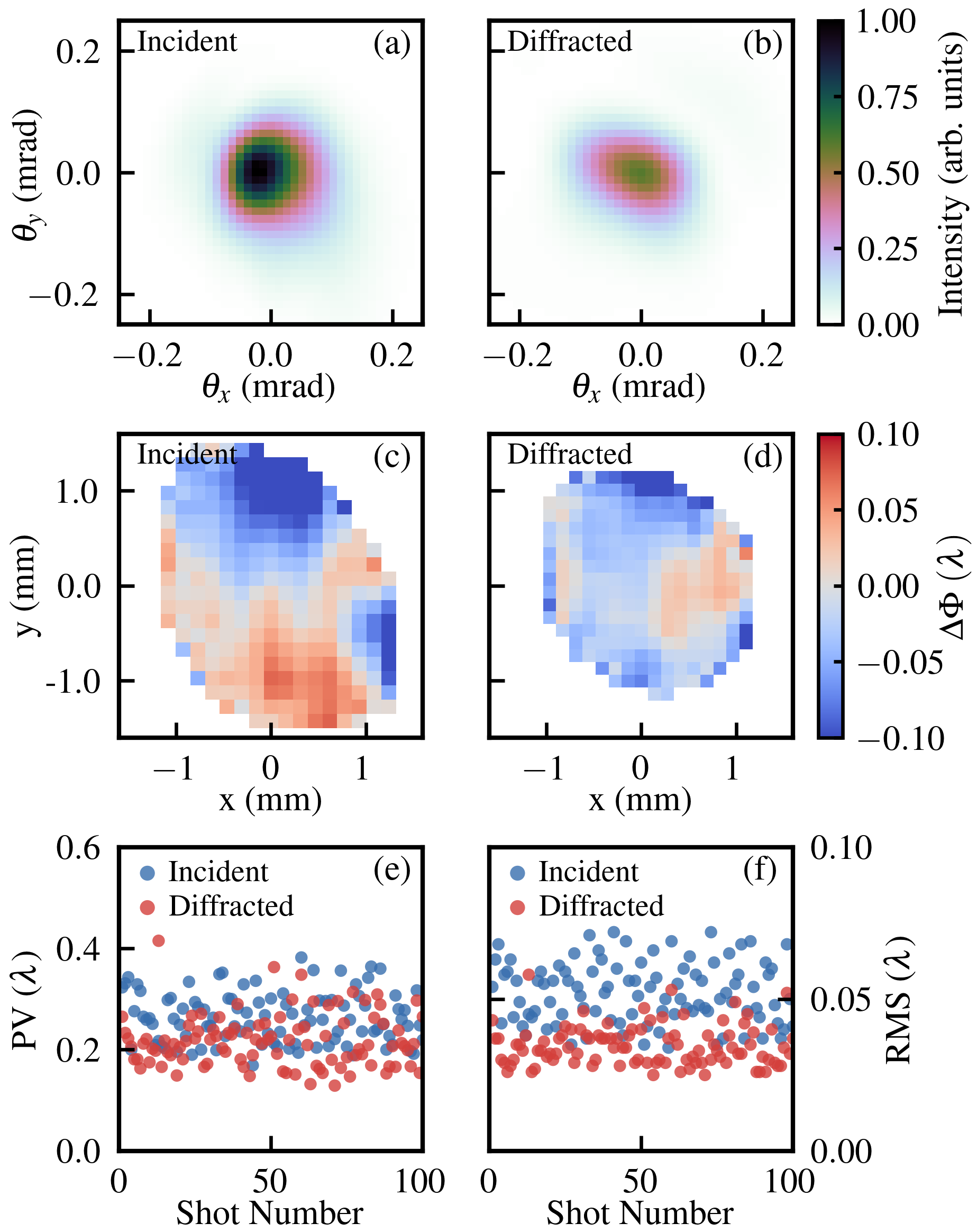}
    \caption{(a, b) Focal spots; (c, d) wavefront; (e) peak-to-valley (PV); and (f) root-mean-square (RMS) wavefront distortion for the incident and first-order diffracted beams. The wavefront was measured using a Shack-Hartmann wavefront sensor. The probe beam was $532\,\mathrm{nm}$ and pulsed. It was cleaned by a spatial filter before entering the grating. The phases are normalized by the probe wavelength.}
    \label{fig:spatial_quality}
\end{figure}

Figure \ref{fig:spatial_quality} shows the spatial quality of the first-order diffracted probe. We imaged the focal spots of the incident and diffracted beams on a camera using a $f = 250\,\mathrm{mm}$ lens. The wavefront was measured using a Shack-Hartmann wavefront sensor (Thorlabs WFS40-7AR). Results in Fig. $\ref{fig:spatial_quality}$(a-d) demonstrate that the gas grating preserves the beam's focusing characteristics without introducing significant aberrations or wavefront distortions. Notably, the peak-to-valley (PV) and root-mean-square (RMS) wavefront distortions shown in panels (e) and (f) indicate that the spatial quality actually improves slightly after diffraction. The average PV and RMS phase errors decrease from the incident to diffracted beam due to the inherent spatial cleaning effect of the gratings. 

\section{Discussion}\label{sec:discussion}
We have built gas gratings that can diffract probe lasers of various wavelengths and pulse durations with extremely high efficiency. These results suggest that the gas gratings can readily be used to manipulate lasers at fluences that would otherwise damage typical solid optics. Compared to plasma gratings and other plasma optics that feature similar advantages in damage thresholds and debris resistance \cite{sheng2003plasma, wu2005manipulating, suntsov2009femtosecond, shi2011generation, durand2012dynamics, lehmann2016transient, peng2019nonlinear,edwards2022plasma, edwards2023control, riconda2023plasma, vieux2023role, edwards2024greater, wu2025spatiotemporal}, gas gratings exist for much longer timescales (hundreds of nanoseconds instead of picoseconds for plasma gratings \cite{edwards2023control}), making them suitable for manipulating long-pulse lasers typically used in high-energy-density (HED) studies and inertial confinement fusion (ICF). Furthermore, gases are generally much more controllable than plasmas. These unique advantages make gas gratings a promising solution to the final optics problem in inertial fusion energy (IFE), where the final optical elements must operate reliably and continuously under extreme radiation and neutron flux from target implosions. 

To summarize the important parameters for constructing a gas grating, we first consider the gas index modulation, which according to the theory \cite{michel2024photochemically} scales as:
\begin{equation}
    n_1\,\sim\,\delta n\,\propto\,\frac{\rho_0}{\gamma}\frac{\Delta T}{T_0},
    \label{eq:dn}
\end{equation}
where $\delta n$ is the total index change (i.e., $n_1$ is the amplitude of the Fourier component of $\delta n$ corresponding to the spatial scale $\Lambda$, the grating period), $\rho_0$ is the background gas density (assuming $n - 1\propto\rho$), $\gamma$ is the heat capacity ratio, $T_0$ is the initial temperature, and $\Delta T$ is the temperature change. In addition,
\begin{equation}
    \Delta T \approx \frac{U}{c_V\,\rho_0},
    \label{eq:dT}
\end{equation}
where $c_V$ is the average heat capacity of the gas and $U$ is the total deposited energy density. Assuming that the imprint fluence can be adjusted to fully deplete the available ozone, since the dissociation of every ozone molecule releases a fixed amount of energy, $U \propto n_A$ where $n_A$ is the absorber (ozone) number density. Combining \eqref{eq:dn} and \eqref{eq:dT} then gives:
\begin{equation}
    n_1 D \propto \frac{n_A\,G\,D}{T_0},
    \label{eq:prod}
\end{equation}
where $G$ is a gas-composition-dependent constant that incorporates several factors: the average gas refractive index and its sensitivity on the density change, the efficiency of the energy deposition process due to various reactions, and other relevant properties of the gas. Equation (\ref{eq:prod}) thus summarizes the four parameters that can be tuned for optimal diffraction, namely the gas composition, absorber density, grating thickness, and initial temperature, provided sufficient imprint fluence and appropriate probe timing. In addition to the constraint imposed above, the grating period needs to be chosen properly such that the angular bandwidth is sufficiently large to efficiently diffract the probe under its entire angular spectrum, while not so large as to support higher-order diffraction.

In conclusion, we have experimentally created gas-phase volume diffraction gratings, achieving high diffraction efficiencies for probe lasers with various wavelengths and pulse durations. We systematically characterized the dynamics and optical properties of these photochemically-induced gas gratings under various conditions. These results can assist optimization of parameters, including imprint fluence, beam timing, and gas composition, in real applications. By demonstrating the critical role of $\mathrm{CO_2}$ in enhancing the index modulation and confirming spatial quality and long-term operational stability, we show that gas gratings provide a promising solution for handling the kilojoule-to-megajoule class beams as the final optics in inertial fusion energy. Furthermore, the validation of the theory \cite{michel2024photochemically} and numerical simulations \cite{oudin2025piafs} allows for the design of gas optics with more complicated structures, for instance a gas-phase diffractive lens \cite{singh2025holographic}.

\medskip \noindent
{\bf Funding.} This work was partially supported by NNSA Grant DE-NA0004130, DOE Grant DE-SC0025497, NSF Grants PHY-2308641, PHY-2206711, PHY-2512131, the Lawrence Livermore National Laboratory LDRD program (24-ERD-001), and by the Gordon and Betty Moore Foundation, Grant DOI 10.37807/GBMF12255. Prepared by LLNL under Contract DE-AC52-07NA27344.

\medskip \noindent
{\bf Disclosures.} Work at Stanford has been supported in part by Xcimer Energy LLC.

\medskip \noindent
{\bf Data availability.} Data underlying the results presented in this paper are not publicly available at this time but may be obtained from the authors upon reasonable request.

\bibliographystyle{apsrev4-2}
\bibliography{references}

\end{document}